\documentclass{elsart3}

\usepackage{graphicx}

\usepackage{amssymb}

\begin{document}

\begin{frontmatter}



\title{Building CMS Pixel Barrel Detectur Modules}

\author[psi]{S.~K\"onig\thanksref{corr}}
\author[psi,uzh]{Ch.~H\"ormann}
\author[psi]{R.~Horisberger}
\author[eth]{B.~Meier}
\author[psi]{T.~Rohe}
\author[eth]{S.~Streuli}
\author[psi,eth]{R.~Weber}
\author[psi]{H.Chr.~K\"astli}
\author[psi]{W.~Erdmann}

\address[psi]{Paul Scherrer Institut, 5232 Villigen PSI, Switzerland}
\address[uzh]{Physik-Institut der Universit\"at Z\"urich, 8057 Z\"urich, Switzerland}
\address[eth]{Institut f\"ur Teilchenphysik, ETH~Z\"urich, 8093 Z\"urich, Switzerland}

\thanks[corr]{Corresponding author; e-mail: stefan.koenig@psi.ch}


\begin{abstract}
For the barrel part of the CMS pixel tracker about 800 silicon pixel detector modules are required. The modules are bump bonded, assembled and tested at the Paul Scherrer Institute. This article describes the experience acquired during the assembly of the first $\sim$200 modules.
\end{abstract}

\begin{keyword}
CMS \sep pixel detector \sep module assembly

\end{keyword}
\end{frontmatter}



\section{The CMS Pixel Barrel Modules}
\label{s1}

{
The barrel part of the CMS pixel detector consists of about 800 detector modules that are mounted in three layers on a structure of thin-walled aluminum pipes and carbon fibre crosspieces~\cite{CMSTDR}. While the majority of the modules (672) are full modules as seen in fig.1 on the right, the edges of the six half-shells are equipped with 16 half-modules each (96 in total, see fig.1 on the left). A half-module has 8 readout chips and is half the size of a full module.

\begin{figure}
\begin{center}
\includegraphics [height = 75mm, width=1.0\columnwidth, keepaspectratio]{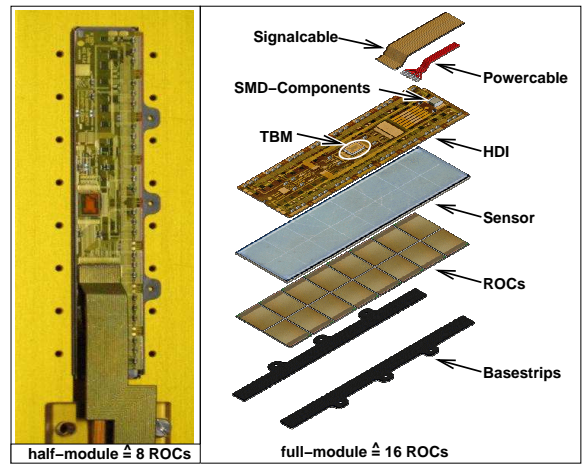}
\caption{Exploded view of a barrel pixel detector full module (right) and picture of an assembled half module (left) }
\end{center}
\end{figure}

As seen on fig.1 a module is composed of the following components:
\begin{itemize}
\item 250~$\mu$m thick silicon nitride basestrips
\item 8 to 16 readout chips (ROCs) with 52x80 pixels of size 150x100~$\mu$m$^2$
\item Sensor made from 285~$\mu$m thick DOFZ-silicon.
\item High Density Interconnect (HDI), a flexible low mass 3 layer PCB (flex print)
\item Token Bit Manager chip (TBM) controling the readout of the ROCs
\item Signal \& Power~cable
\end{itemize}

A completed full-module has the dimensions 66.6x26.0~mm$^2$, weight 2.2~g plus up to 1.3~g for cables, and consumes 2~W of power.
More details can be found in~\cite{hchr_vertex}.
}



\section{Assembly Procedure}
\label{s2}

{

The assembly procedure of a CMS barrel module divides in four parts (see fig 2):

\begin{enumerate}

\item assembly of the HDI

\item processing of the ROC- and sensor- wafers

\item bump bonding of the raw module

\item final assembly of the module

\end{enumerate}

}


\subsection{HDI Assembly}
\label{s3}

{
The flex print is delivered pretested and equipped with all passive electrical components by the vendor. Afterwards the power and signal-cable and the TBM-chip are attached and connected with aluminum wirebonds at PSI. The assembled HDI is then tested for the functionality of the TBM and any shortcuts or broken traces in the power line or token passage.
Up to 16 HDIs can be assembled per working day.

In the early stage of the production the bondability of the HDIs after several month of storage at room conditions was not satisfactory. It was then improved by a mechanical cleaning step with a soft rubber. Unfortunedly this method was likely to create shortcuts between adjacend bondpads and caused a considerable amount of extra work in testing and debugging of assemled HDIs and fabricated modules (23~\% of HDIs not useable, 22~\% of modules needed repairs). The storage of the flexes in a dry nitrogen atmosphere introduced directly after the first problems were discovered eliminated the need for cleaning the bondpads and led to an excelent yield of the assembled HDIs. The fraction of failing HDIs was reduced to less than $<$2\%.

}


The different parts are addressed in the following subsections.

\begin{figure}
\begin{center}
\includegraphics [height = 90mm, width=1.0\columnwidth, keepaspectratio]{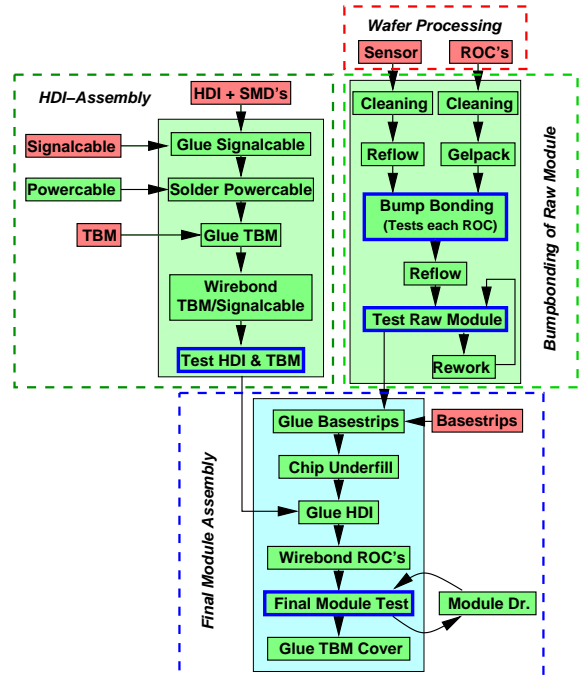}
\caption{Flowchart showing the assembly procedure}
\end{center}
\end{figure}


\subsection{Processing of ROCs and Wafers}
\label{s4}

{
The sensor and ROC wafers have to undergo certain processing and manipulation steps before they can be used for the module production:

First of all an acceptance test of sensors and ROCs is conducted on the wafer.
After 72 ROC-wafers tested the overall yield is about 70\%. The acceptance-tests on sensor wafers reflect just the results of the vendor.
Then follows a composition of photolithographic steps, UBM\footnote{Under Bump Metal - for ROC wafers only, sensor wafers have UBM by the vendor} sputtering and indium evaporation and a final liftoff step to prepare the wafers for the subsequent bump bonding process~\cite{BUMPBOND}
The ROC wafers are then thinned down to 170~$\mu$m and ROC and sensor-wafers are diced and picked based on the results of the acceptance tests.
Finally the raw devices are tested again. In the case of the ROCs this is done as part of the bump bonding process.

Only chips with less than 1\% statistically distributed dead or noisy pixels and without pixel masking defects are used in the following assembly steps.
A sensor is discarded when either the dark current is above 2~$\mu$A at 150~V and room temperature or the ratio of the dark currents measured at 100~V and 150~V is greater than two.

The sensors are then cleaned and undergo the reflow process where the deposited indium is melted up to prepare the indium bumps needed for the later bump bonding. After another optical inspection the sensors are stored unter a dry nitrogen atmosphere untill they are used.

Initial problems during the liftoff-process leading to very work intensive procedures could be solved by modifying the fotolithographic masks. Now the openings in the fotoresist are distributed over the whole surface of the wafer. This reduced both the duration and complexity of the liftoff procedures and reduced the use of the ultrasonic bath that caused partial removing of the deposited indium bumps on several sensors to an absolute minimum. Another problem with corrosion like effects on the aluminum bondpads of the ROC chips was solved by consequently renewing the liftoff solvend after each liftoff step, especially before the use of the ultrasonic bath.
}



\subsection{Raw Module Assembly}
\label{s5}

{
The sandwich of sensor and the 8 to 16 ROC chips needed to build a module is called 'raw module'. Building it involves the following steps:

First the set of ROCs is cleaned and arranged with the help of a special jig that ensures their defined positions on the gelpacks which are then placed on the bump bonding machine. Sensor and ROCs are then bump bonded using the automatic bump bonding machine developed at PSI~\cite{BUMPBOND} (see fig. 3). The individual ROC chips are tested during the bump-bonding process to ensure that only electrically working chips are bump-bonded to the sensor. To form a mechanical and electrical connection between the indium bumps of sensor and ROCs the module is reflowed in the reflow oven.
Afterwards the completed raw module is tested again.

\begin{figure}
\begin{center}
\includegraphics [height = 60mm, width=1.0\columnwidth, keepaspectratio]{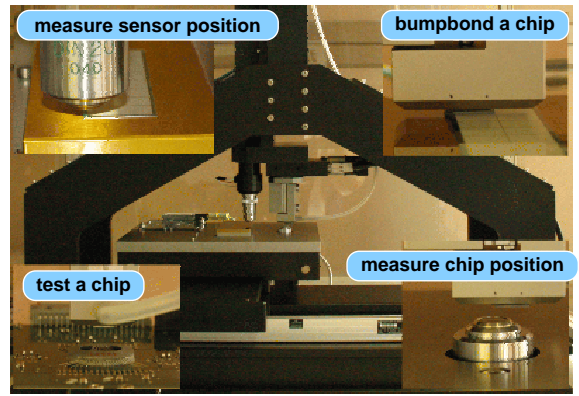}
\caption{Automatic bump bonding machine at PSI }
\end{center}
\end{figure}

\begin{figure}
\begin{center}
\includegraphics [height = 80mm, width=1.0\columnwidth, keepaspectratio]{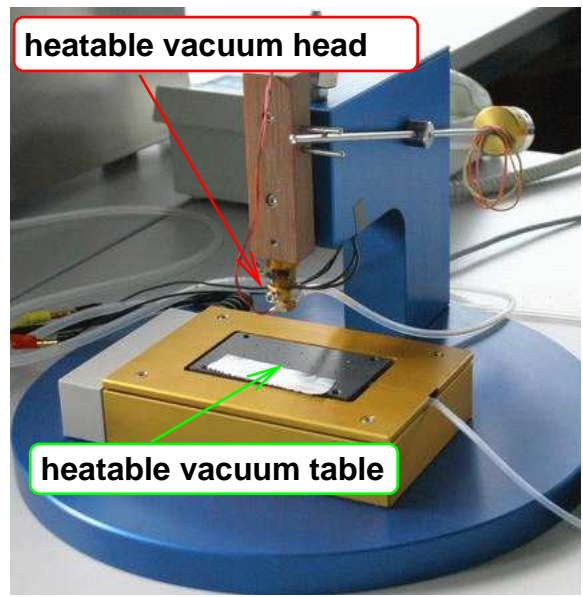}
\caption{Reworking jig at PSI}
\end{center}
\end{figure}

The so called 'raw test' gives a very fast feedback on the bump bonding quality and was upgraded several times to enhance its abilities. It finally measures the ROC power consumption, does a threshold scan and measures the level of the analog output. In addition to that it does an IV curve measurement on the sensor and makes a bump yield test. Based on the result of the test the modules are released for further assembly or undergo a reworking step to remove defective chips or improve the bump yield if possible and usefull. Raw modules drawing a high current ($>$15~$\mu$A @ 200V) are not repaired and are not used further on. The ability to rework single chips of a bare module with a special reworking jig (see fig. 4) has constantly improved during the module production. The rework quality is either excellent (less than 12 bump defects per chip) or very poor ($\sim$1000~defects per ROC). The overall chip replacement yield is about 85\% now (60 of the 71 replaced chips have been good) and even full modules (16 chips) were reworked with almost perfect results (see fig. 5).

\begin{figure}
\begin{center}
\includegraphics [height = 60mm, width=1.0\columnwidth, keepaspectratio]{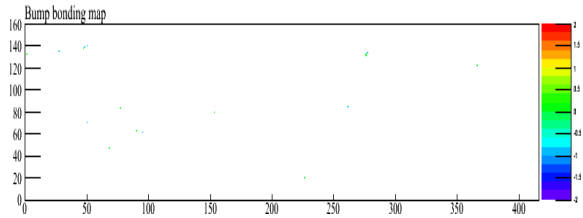}
\caption{Bump yield plot of a completely reworked module (16 chips reworked) only very few bump defects (colored dots) could be found}
\end{center}
\end{figure}

}


\subsection{Complete Module Assembly}
\label{s6}

{
When the raw module has passed the last test, the construction of the module is finalised on the PSI module assembly line (see fig. 6) in the following steps:

 First the base strips are glued to the ROC side of the raw module. After the glue has cured and a small amount of glue has been applied between the edge of the sensor and the ROCs to reinforce the connection near the wirebonding area the pre-assembled HDI is glued on the sensor side of the raw module. After the electrical connection between the flex and the ROCs is formed with wire bonds the module assembly is complete.
Each glueing step is done on a separate jig that ensures the exact placement of the parts and keeps them in place with vacuum until the glue has cured.

Instead of using a glue disperser to apply the desired small quantities of glue, the adhesive is applied with a stamp that is lowered into a glue bath and matches the form of the glueing area. This method is a very easy but well defined way to apply the glue exactly where it is needed.

In order to apply the very small amount of glue in between the ROCs and the sensor, the stamp is replaced with Kapton${\textcircled {\scriptsize R}}$~\cite{KAPTON} flaps that bring the glue in the opening between the sensor and the ROCs.

The assembly procedure has proven itself to be very reliable and fault proof during the production so far and has not been changed even after more than 200 modules build.
The number of modules that after assembly are worse than before is only about 8, which is less than 4\% of the modules produced.

At the end the completed module undergoes the final testing. This includes thermocycling between room temperature and -20~{\ensuremath{^\circ}}C. The fully automatic test setup (see fig. 7) can handle up to 4 modules in parallel and up to 2 testloads per working day.

The modules are then graded according to their test results. The major criterias are the following:

Grade A modules are excellent with less than 1~permill defectice or unbonded pixels per chip and sensor dark currents of less than 2~$\mu$A at 17~{\ensuremath{^\circ}}C and 150~V.

Grade B modules are useable modules with less than 4~percent defectice or unbonded pixels per chip and sensor dark currents up to than 10~$\mu$A at 17~{\ensuremath{^\circ}}C and 150~V.

All other modules are graded as C if they cannot be repaired by the 'module doctor', which is the case for sensor defects (high dark current) or chip defects. If only a bond has loosened or a short on the HDI has developed or even the TBM ceases to function the module will be repaired and tested again.
The 'module doctor' was introduced to understand and possibly repair the HDI-induced module defects and has proven to be very usefull in the first phase of the production, where many modules needed to be repaired. Due to the improved quality of the assembled HDIs the 'module doctor'-station was used only very frequent during the production of the last 100 modules.

\begin{figure}
\begin{center}
\includegraphics [height = 80mm, width=1.0\columnwidth, keepaspectratio]{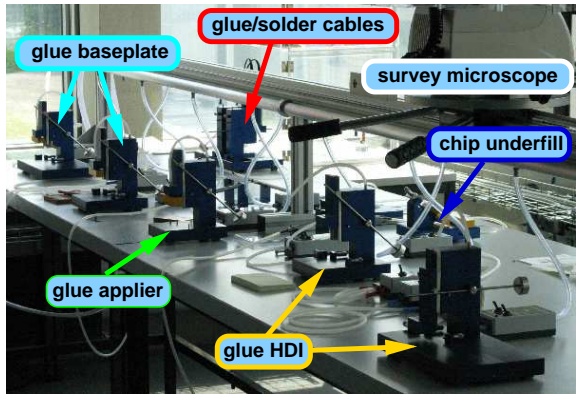}
\caption{Assembly line at PSI}
\end{center}
\end{figure}

The module assembly procedure is capable of a constant rate of 4 full and 2 half modules per working day. In parallel to that 16 HDIs can be assembled together with the processing of ROCs and sensors for the bump bonding. The manpower required to maintain this rate is about 5 FTEs\footnote{Full Time Equivalents} including the final testing of the modules.

More Details about the module assembly can be found in \cite{BUILD}.

\begin{figure}
\begin{center}
\includegraphics [height = 80mm, width=1.0\columnwidth, keepaspectratio]{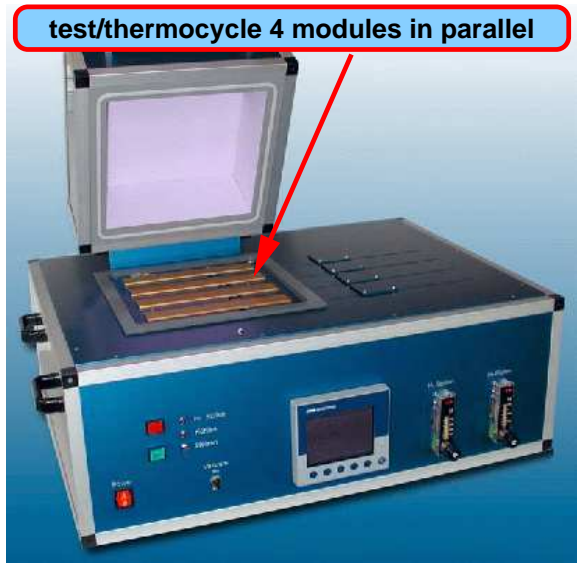}
\caption{Setup for final module test including thermocycling }
\end{center}
\end{figure}

}



\section{Production Result}
\label{s7}

{
Up to end of November 2007, 286~Modules were bump bonded, but 30 modules were in different stages of production and testing at this time. Out of the 256 completed moduls 210 (82\%) modules are graded A(177) or B(33) (59 after repair or rework). Half of the 46 modules with grade C failed in the state of a raw module.
About half of the modules were lost due to problems and errors in the startup of the production. The others are rejected due to sensor problems.

}




\begin{thebibliography}{14}





\bibitem[1]{CMSTDR} The CMS Collaboration. CMS Tracker. Technical Design Report LHCC 98-6, CERN, Geneva, Switzerland, 1998.
\bibitem[2]{hchr_vertex} CMS Barrel Pixel detector Overview. H.C. Kaestli these proceedings.
\bibitem[3]{PSI46} H.C. Kaestli. Design and performance of the CMS pixel detector readout chip. NIM A 565(2006) 188-194
\bibitem[4]{SENSOR} T.Rohe et al.Fluence Dependence of Charge Collection of irradiated Pixel Sensors. arXiv:physics/0411214 v3 4 Jan 2005
\bibitem[5]{KAPTON} Kapton${\textcircled {\scriptsize R}}$ is a registrated trademark of DuPont for its polyimide film
\bibitem[6]{BUMPBOND} Ch. Br\"onnimann et al. Development of an Indium Bump Bond Process for Silicon Pixel Detectors at PSI. These proceedings.
\bibitem[7]{BUILD} S. K\"onig. Assembly of the CMS pixel barrel modules. NIM A 565(2006) 62-66

\end{thebibliography}
\end{document}